\documentclass[sigconf,authorversion,nonacm]{acmart}
\usepackage{multirow}
\usepackage{graphicx}
\usepackage{tabularx} 
\usepackage{booktabs} 
\usepackage{array}
\usepackage{geometry}
\usepackage{adjustbox}
\usepackage{enumitem}
\usepackage{pdfpages}

\setlist[itemize]{leftmargin=*}

\newcommand{\sys}{{REVA}{}}

\AtBeginDocument{%
  }

\setcopyright{acmlicensed}
\copyrightyear{2018}
\acmYear{2018}
\acmDOI{XXXXXXX.XXXXXXX}

\acmConference[Conference acronym 'XX]{Make sure to enter the correct
  conference title from your rights confirmation emai}{June 03--05,
  2018}{Woodstock, NY}
\acmISBN{978-1-4503-XXXX-X/18/06}
\begin{document}

\title{\sys{}: Supporting LLM-Generated Programming Feedback Validation at Scale Through User Attention-based Adaptation}

\author{Xiaohang Tang}
\affiliation{%
  \institution{Virginia Tech}
  \city{Blacksburg}
  \state{Virginia}
  \country{USA}
}
\email{xiaohangtang@vt.edu}

\author{Sam Wong}
\affiliation{%
  \institution{University of Washington}
  \city{Seattle}
  \state{Washington}
  \country{USA}
}
\email{samw627@uw.edu}

\author{Zicheng He}
\affiliation{%
  \institution{ University of Virginia}
  \city{Charlottesville}
  \state{Virginia}
  \country{USA}
}
\email{bgc4bx@virginia.edu}

\author{Yalong Yang}
\affiliation{%
  \institution{Georgia Institute of Technology}
  \city{Atlanta}
  \state{Georgia}
  \country{USA}
}
\email{yalong.yang@gatech.edu}

\author{Yan Chen}
\affiliation{%
  \institution{Virginia Tech}
  \city{Blacksburg}
  \state{Virginia}
  \country{USA}
}
\email{ych@vt.edu}

\renewcommand{\shortauthors}{Xiaohang Tang, Sam Wong, Zicheng He, Yalong Yang, and Yan Chen}
\begin{teaserfigure}
  \includegraphics[width=\textwidth]{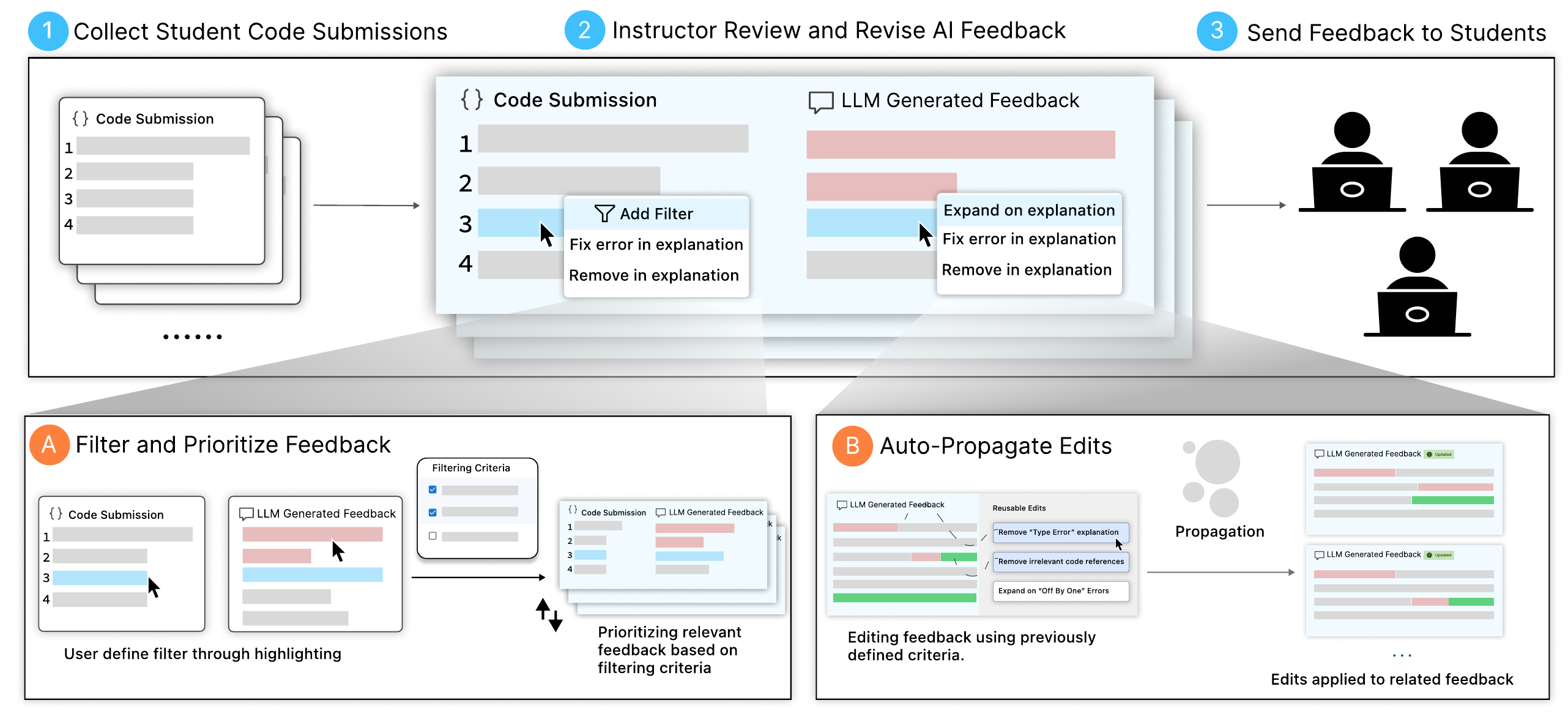}
  \caption{The workflow of \sys{}. \textmd{There are three steps to reviewing and providing AI-generated programming feedback at scale using \sys{}. (Step 1) Collect Student Code Submissions where instructors gather submissions from hundreds of students to a coding exercise. (Step 2) Instructor Review and Revise AI Feedback, the core innovation of \sys{}, where instructors efficiently validate feedback through two key features: defining filters by highlighting relevant content \textcircled{\raisebox{-0.5pt}{A}} to intelligently reorder the review queue and minimize context switching, and making revisions that automatically propagate to semantically similar instances \textcircled{\raisebox{-0.5pt}{B}}, significantly reducing repetitive work. (Step 3) Send Feedback to Students once the review and revision process is complete.}}
  \Description{}
  \label{fig:teaser}
\end{teaserfigure}
\renewcommand{\shorttitle}{\sys{}:  Supporting LLM-Generated Programming Feedback Validation at Scale}
\begin{abstract}
This paper introduces \sys{}, a human-AI system that expedites instructor review of voluminous AI-generated programming feedback by sequencing submissions to minimize cognitive context shifts and propagating instructor-driven revisions across semantically similar instances.
\sys{} introduces a novel approach to human-AI collaboration in educational feedback by adaptively learning from instructors' attention in the review and revision process to continuously improve the feedback validation process. \sys{}'s usefulness and effectiveness in improving feedback quality and the overall feedback review process were evaluated through a within-subjects lab study with 12 participants.

\end{abstract}
\maketitle

\section{Introduction}
When learning to program, personalized feedback is among the most effective ways to help students bridge the gap between their current understanding and their desired learning goals~\cite{butler1995feedback,hattie2012visible}. According to feedback model~\cite{hattie2007power}, effective personalized feedback consists of feed-up (issues), feed-forward, and feedback, which of all should be personalized. However, as class sizes grow, providing timely, personalized feedback becomes increasingly challenging due to the significant time and cognitive demands required to thoroughly examine each student's submission.

While Large Language Models (LLMs) offer the potential to generate personalized feedback at scale~\cite{macneil2022generating,yadav2023contextualizing,xu2024critique}, their reliability remains a critical challenge. LLM-generated responses can exhibit inaccuracies~\cite{Bender2021OnTD}, hallucinations~\cite{huang2023surveyhallucinationlargelanguage}, or fail to align with course-specific learning objectives and pedagogical approaches~\cite{Sridhar2023HarnessingLI, sonkar2024pedagogical,jacobsen2025ai}. 
Consequently, instructor validation is a necessary step to guarantee the quality of feedback provided to students. 
Yet, this validation process itself imposes a significant cognitive burden, demanding instructors evaluate large volumes of code-feedback pairs.
Addressing this cognitive bottleneck is the core motivation for our research, which seeks to enhance instructor efficiency and performance in reviewing AI-generated feedback.

The process of reviewing AI-generated feedback can be conceptualized as a sequence of cognitive tasks, each requiring instructors to mentally shift between different programming concepts, error patterns, pre-existing knowledge, and pedagogical approaches. Research in psychology demonstrates that such frequent context switching imposes significant increases in mental fatigue and reductions in efficiency~\cite{monsell2003task,wylie2000task}. 
Yet, in educational settings, the review process contains inherent structural patterns: although presented as individual items, code-feedback pairs often cluster around similar code concepts or error patterns, creating opportunities for increased cognitive continuity between review tasks. 
To capitalize on these opportunities and mitigate the cognitive burden, we propose leveraging \textbf{user attention-based adaptation}, an approach where the system dynamically learns from indicators of the instructor's focus (their attention) to intelligently adapt the review workflow.

While user attention-based adaptation offers a promising path to harness these structural patterns for improved cognitive continuity, existing tools generally fall short in this regard.
They often fail to leverage semantic repetitions inherent in the feedback review process, instead treating each AI-generated item as an isolated unit rather than part of a connected workflow. 
For instance, tools like OverCode~\cite{glassman2015overcode} help instructors create cluster-based feedback across student solutions but do not address feedback review. Gero et al.~\cite{SensemakingLLMGero2025} explored supporting the sense making of LLM-generated content by helping users compare and contrast different outputs, but did not address the cognitive continuity challenges in sequential review tasks. Template-based systems such as LA Cockpit~\cite{karademir2024LACockpit} struggle with the dynamic nature of student submissions. Although approaches like FIXPROPAGATOR~\cite{head2017writing} and GPT4Hints-GPT3.5Val~\cite{TutorStyleProgrammingFeedbackPhung} uses AI to propagate code fixes or validate feedback, they did not address the cognitive challenges instructors face when reviewing numerous code-feedback pairs with substantial semantic repetition but literary variation.
The critical question remains: \textbf{\textit{How can we design a system that leverages the natural semantic structure of programming feedback to support cognitive continuity during the review process, thereby reducing cognitive switching costs and improving review efficiency?}}

In this paper, we present \sys{}\footnote{\sys{} stands for feedback \textbf{REV}iew and \textbf{REV}ision at scale through user \textbf{A}ttention-based \textbf{A}daptation
}, \textbf{a human-AI system} that expedites instructor review of voluminous AI-generated programming feedback.
\sys{} operationalizes the concept of user attention-based adaptation through two key mechanisms: \textit{adaptive content sequencing} and \textit{revision propagation}. Our insight is that elements capturing instructors' attention likely contain the highest information value, and by using these attention patterns to sequence semantically similar feedback instances, we can minimize cognitive switching costs while automatically propagating revisions across related instances.

Figure~\ref{fig:teaser} depicts the workflow of \sys{}. Once student code submissions are collected (Step 1), instructors use an LLM to generate feedback for each one of them through our component-based feedback generation prompt~\cite{hattie2007power} which structures feedback according to established learning models. The core of \sys{} (Step 2) is where instructors review and revise AI-generated feedback. During this critical step, instructors can define filters by highlighting relevant content in code or feedback (A) to prioritize which submissions to review next, minimizing cognitive load from context switching between semantically different submissions. They can also make targeted revisions that \sys{} automatically propagates to semantically similar feedback instances (B), significantly reducing repetitive editing work. After the review and revision process is complete, instructors can send the refined feedback to students (Step 3). 

The significance of \sys{} lies in how it transforms the traditionally labor-intensive process of reviewing AI-generated programming feedback into an adaptive, instructor-centered workflow. By intelligently leveraging attention patterns and revision behaviors, \sys{} lets instructors to efficiently review large volumes of feedback that would otherwise be cognitively overwhelming, ultimately making personalized programming education feasible at scale. 

After conducting a formative study ($N=20$) and identifying three key patterns that support \sys{}'s design, we conducted a within-subject study with 12 programming instructors experienced in grading and providing feedback to evaluate \sys{}'s usability and effectiveness. Participants reviewed 475 AI-generated feedback using both \sys{} (with attention-based filtering and revision propagation) and a baseline version without these features. 
Results showed that sys{} helps instructors make significantly more revisions, and produce significantly better feedback regarding the precision and recall of misconception coverage and overall feedback quality. We also observed that \sys{}'s semantic filters and revision propagation helped participants react significantly faster while switching contexts and kept them engaged with feedback validation across time.

Our research makes the following contributions:
\begin{itemize}
    \item Design implications for supporting users in reviewing large amounts of AI-generated code feedback using users' attention.    
    \item A new approach that collects and leverages users' attention and intention seamlessly in feedback reviewing and refining interactions to adapt the presentation and order of review, and propagate the refinements to upcoming feedback in a controllable way.
    \item An novel system, \sys{}, that implements this approach with a mixed-initiative interface for effective human-AI collaboration in creating, reviewing, and revising code feedback at scale.
\end{itemize}

\section{Related Work}

Our work draws inspiration primarily from four areas: task sequencing and cognitive transitions, personalized feedback at scale, document review and LLM output evaluation, and instructors' roles in classroom activities. In this section, we review the relevant literature and identify the key needs and challenges in these domains.

\subsection{Task Sequencing and Cognitive Transitions}
The challenge of reviewing large volumes of AI-generated programming feedback shares structural similarities with research on task sequencing and cognitive transitions. Studies in cognitive psychology have established that switching between tasks with different mental frameworks imposes significant cognitive costs~\cite{monsell2003task,wylie2000task}. These switching costs manifest as increased time, reduced accuracy, and higher mental fatigue as individuals must reconfigure their mental models for each new context~\cite{borst2015makes,bailey2008understanding}.

From our observations and prior work, when instructors review programming feedback, they face repeated cognitive transitions between different code error patterns, and feedback. Each new submission requires instructors to build a comprehensive mental state that encompasses understanding the student's approach, identifying misconceptions, and formulating appropriate guidance. This mental state is particularly vulnerable to disruption when transitioning between conceptually dissimilar submissions.
Although individual code-feedback pair may vary, they often share contextual similarities, such as addressing the same code error or similar programming concepts, leading to substantial semantic (not literary) repetition. These similarities represent opportunities for cognitive continuity that existing systems fail to leverage. 

Our work extends these principles of cognitive task sequencing to programming feedback review. By treating each feedback instance as a cognitive task with specific properties (users' attention, code-feedback semantics), we can develop an adaptive system that sequences the review process to minimize disruptive transitions. 

\subsection{Personalized Feedback at Scale}
Providing timely and effective feedback is crucial for enhancing student learning. Feedback aims to ``reduce discrepancies between current understandings/performance and a desired goal'' and has one of the most powerful influences on student learning~\cite{hattie2007power}. Effective feedback must be accurate, timely, and tailored to students' needs~\cite{chen2020edcode}. Delayed or poor feedback, such as negative or incorrect feedback, can lead to worse performance than no feedback at all.

Creating real-time, personalized feedback is particularly challenging due to the multifaceted data involved.
In our context of group coding exercises, this includes both coding and natural language (e.g., discussion messages) and group dynamics (e.g., different roles). Prior research shows the potential for instructors to analyze students' code~\cite{zhang2024CFlow, glassman2015overcode} and oversee students' collaborative learning behaviors~\cite{yang2023pair, tang2024vizgroup} at scale.
In practice, teachers using LA tools typically focus on individual student performance or the entire class as a whole~\cite{wu2024impact}. As a result, their responses are often limited to either individual or whole-class scaffolding, without fully leveraging the data and setting of collaborative coding exercises. This leaves significant potential untapped, as the primary goal of group discussions is to utilize peer interactions to reduce teachers' workload both for understanding the students' issues and provide peer support~\cite{wang2021puzzleme}.
Therefore, there is a clear need to bridge this gap by utilizing the group setting and co-created data to enhance feedback mechanisms effectively. 

On the other hand, the recent advances in AI have led to various approaches in feedback creation for code understanding~\cite{nam2024using} and auto feedback generation~\cite{gabbay2024combining}. While promising, studies have shown that the quality and reliability of auto-generated feedback can be as low as 50\% accurate in assessing the correct mistakes students made in programming assignments~\cite{estevez2024evaluation}. Additionally, students often find AI-generated feedback hard to understand, unfriendly, and requiring manual validation~\cite{nguyen2024beginning}. Instructors also worry that fully automated feedback generation could lead to the misuse of the responses and are concerned about the general relevance and alignment with learning objectives.
This indicates a need for instructors to review feedback before sending it out.

\subsection{Documents and LLM Output Evaluation}

Document review is a cognitively demanding task, especially when dealing with large volumes of text~\cite{pirolli1999information}.
Traditional document visualization and comparison tools like OverCode~\cite{glassman2015overcode} and Examplore~\cite{glassman2018visualizing} focus on identifying similarities and differences between documents to determine which are ``better'' in some dimension. However, our work addresses a fundamentally different challenge: evaluating whether AI-generated feedback aligns with instructors' pedagogical needs and accurately addresses code issues at scale.
When instructors review AI-generated feedback, they must rapidly determine whether each feedback instance is relevant to the corresponding code submission, accurate in its assessment, and pedagogically appropriate. This evaluation requires a distinct cognitive process compared to document comparison, as instructors must maintain context across both code and feedback while making judgments about educational value. Existing tools fail to support this specific review pattern~\cite{SensemakingLLMGero2025}, where the goal is not to compare documents but to validate alignment between feedback, code, and learning objectives.

More recently, the HCI community has started exploring ways to support users in evaluating LLM outputs. Tools like ChainForge~\cite{arawjo2024chainforge} and EvalLM~\cite{kim2024evallm} primarily support evaluators in comparing outputs from different prompts to optimize prompt engineering. However, these tools treat evaluation as a means to identify better prompts rather than directly improving the outputs.
Other work like Graphalogue~\cite{jiang2023graphologue} and Sensescape~\cite{suh2023sensecape} transforms long LLM responses into diagrams that connect concepts using a graph structure to enhance the rendering of individual LLM responses. Our work differs by focusing on the efficient review of AI-generated educational feedback at scale. This includes evaluating whether feedback addresses specific code issues, provides targeted guidance for fixing problems, and maintains consistency with course learning objectives.

Our approach draws inspiration from interaction design techniques like programming-by-demonstration~\cite{lieberman2001your,cypher1993watch} and automatic propagation mechanisms~\cite{gulwani2011automating}, which have been explored in other domains but not applied to the unique problem structure of programming feedback review. By adapting these techniques to leverage the semantic repetition inherent in code-feedback pairs, \sys{} enables instructors to efficiently review and refine feedback at a scale that would otherwise be prohibitively demanding.

\subsection{Instructors' Role and Effort in Classroom Activities}
Human instructors play an irreplaceable role in educational settings, particularly in in-class activities~\cite{comas2011learning, ccardak2016increasing}. Unlike AI tools, instructors possess a unique ability to understand and adapt to the complex cognitive and emotional landscapes of their students, tailoring feedback based on each student's progress and challenges. However, while conducting in-class activities offers more opportunities for teacher-student engagement, research shows that the variety of student behaviors during these activities demands significant effort and flexibility from teachers~\cite{comas2011learning}. This includes effectively monitoring student activities~\cite{zhang2023vizprog} and making appropriate decisions in real-time~\cite{van2010scaffolding}.
There is a pressing need for strategies and tools to support instructors in efficiently identifying and acting upon critical issues during exercises, especially at scale. By addressing this gap, we aim to develop approaches that can augment human instructors' capabilities in real-time, enhancing the quality and effectiveness of interactive learning experiences in programming education while preserving the irreplaceable role of human instructors.
\begin{figure*}[h]
    \centering
    \includegraphics[width=1\linewidth]{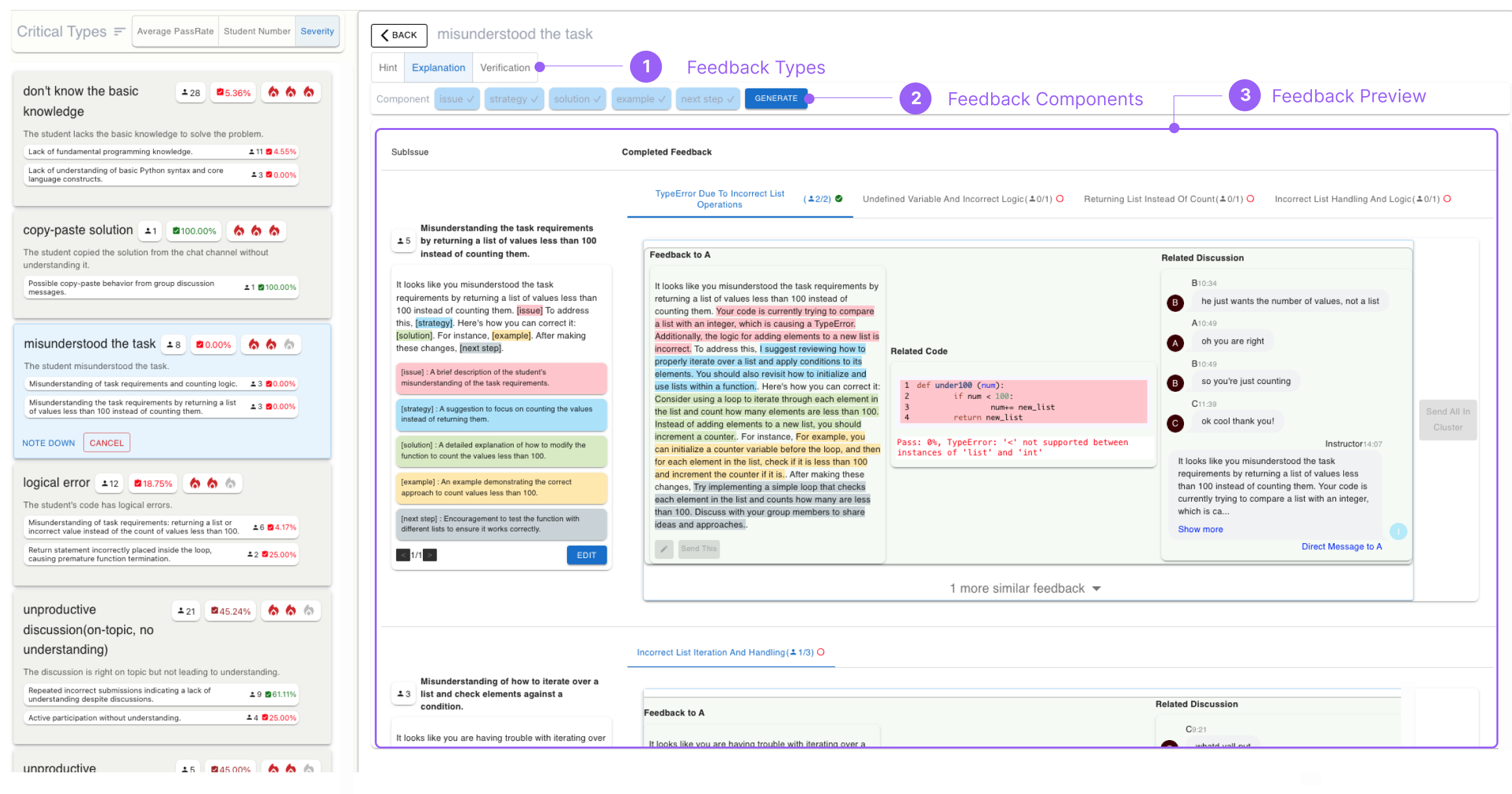}
    \caption{\textmd{User Interface of the probe system for creating and reviewing LLM-generated feedback. (1) Instructor can select different feedback types, which will select a subset of (2) feedback components. (3) Instructors can then generate feedback for preview.}}
    \label{fig:probe}
\end{figure*}
\section{Formative Study}
In order to understand instructors' needs and workflow they use when authoring large number of LLM-generated feedback to students, we developed a design probe that allows instructors to author feedback using LLMs in both a systematic way and at scale. 

\subsection{Design Probe}
The probe groups student code submissions based on common errors and displays the number of students within each group (Figure~\ref{fig:probe}). Drawing on Hattie et al.'s feedback model~\cite{hattie2007power}, each piece of feedback is generated using a template with five components: Issue, Strategy, Solution, Example, and Next Step. Participants can easily toggle each component on or off, depending on their preferences. Each feedback entry includes both textual guidance and the student's code submission, along with their conversation history. To ensure clarity and traceability, we incorporate highlighting-a design paradigms for user interfaces that support the validation of large-scale LLM outputs \cite{Gero2024Sensemaking, kambhamettu2024traceable}, to the feedback. This allows users to identify which parts of the feedback relate to specific elements in the code or conversation—for example, by highlighting corresponding segments in the feedback and the associated evidence. Additionally, feedback groups are further clustered by similarity, enabling instructors to send the same feedback to all students within a given cluster.  Based on prior work~\cite{10.1145/3636555.3636905}, the probe system also employs a custom LLM framework that identifies students with critical issues according to their coding behaviors to help instructors select students to give feedback.

\subsection{Methodology}
We conducted an in-person, between-subject user study—approved by our institution's IRB—to evaluate the design probe's usability and effectiveness in identifying student issues and delivering feedback. We recruited 20 instructors (7 female, 11 male, 2 non-binary) from a four-year university, all with teaching and programming experience, using personal networks, local mailing lists, and snowball sampling. Participants received \$15 USD for reviewing student behavior data from a large programming course. To simulate a real-time teaching environment, they watched a playback of a coding session with 111 students working individually on a Python task, followed by collaborative discussions. Each participant was assigned to one of two conditions: Baseline, a minimal version of the design probe with manual prompt editing and no structured feedback, or the design probe’s full-featured version with structured templates and visual augmentations.
Participants completed three tasks designed to evaluate feedback quality in a live teaching context. In Tasks 1 and 2, they used the critical issue recommendation panel to assist struggling students within two 6-minute intervals. In Task 3, they provided targeted feedback to specific low-performing students. Each 45 minute session included informed consent, a system tutorial, task execution, post-task surveys including a NASA-TLX questionnaire, and a semi-structured interview. Sessions were screen and audio recorded, and participants were encouraged to think aloud throughout.

\subsection{Results and Design Consideration}
\subsubsection{Design probe enables generation of higher quality feedback compared to baseline}

In total, we collected 5,871 feedback messages (5,049 generated, 107 edited, and 715 sent). Two researchers independently annotated the feedback as \textit{incorrect}, \textit{shallow}, or \textit{high-quality}. Component-based generation was found to be an effective method for producing higher-quality feedback. Based on the sampled dataset, participants using design probe generated significantly more high-quality feedback than those in the baseline condition (Probe: $\mu = 80.83\%, \sigma = 0.22$; baseline: $\mu = 45.83\%, \sigma = 0.28$; $p < 0.01$), while also generating less incorrect feedback (Probe: $\mu = 14.16\%, \sigma = 0.15$; baseline: $\mu = 44.17\%, \sigma = 0.32$; $p < 0.05$). Furthermore, participants using design probe sent significantly more high-quality feedback (Probe: $\mu = 80.17\%, \sigma = 0.17$; baseline: $\mu = 46.33\%, \sigma = 0.28$; $p < 0.01$), and less incorrect feedback (Probe: $\mu = 9.17\%, \sigma = 0.15$; baseline: $\mu = 45.00\%, \sigma = 0.31$; $p < 0.01$) in the sampled sent dataset. There was no significant difference in the amount of shallow feedback between the two conditions (Probe: $\mu = 10.67\%, \sigma = 0.17$; baseline: $\mu = 8.67\%, \sigma = 0.14$). We used T-test for numerical feedback data and Fisher's Exact Probability Test for feedback quality transition data. These statistics informed our design of a feedback generation and validation interface.

\subsubsection{Instructors develop implicit mental models to prioritize content during review}
\label{sec: pattern1}
The design probe reveals that instructors rely on their own intuition when prioritizing which feedback needs to be reviewed. Specifically, they develop personal strategies for determining which feedback is worth addressing and how it should be addressed. Instructors often prioritize certain elements of the feedback based on the severity of the issue in the current submission.\textit{“Based on the severity that I choose—if I choose a very high level—sometimes I don’t really look at the next step, because after I read the AI-generated feedback, the next step is sometimes just, like, ‘You can talk with your group members’” (P5)}. Others prioritize the nature of the problem when giving feedback: \textit{“My purpose is to try to help the students—just based on the code. So sometimes I think the solution example is already enough to help students solve the problem. So I will create the next step” (P6).} Furthermore, some participants' intuitive approaches to giving feedback do not necessarily align with the structure of the current system: \textit{“It would’ve been cool if I could just generate feedback for one of the sub-issues, rather than the full thing” (P15).} The diversity of these intuitive approaches informs our design goal of creating an adaptive interface that accounts for users’ attention and priorities (DG1).

\subsubsection{Instructors prefer to review similar programming issues or feedback in the same batch to reduce context switching}
\label{sec: pattern2}
When reviewing large amounts of feedback, instructors aim to minimize the effort required. To that end, they often leverage the system’s clustering feature to review feedback in batches, which helps reduce the cognitive load. For instance, 5 out of 10 instructors found that clustering the feedback streamlined the review process. As P10 noted, clustering “supports handling the problem in chunks,” ultimately aiding in “understanding students’ behaviors and giving feedback quickly.” Similarly, P9 emphasized that clustering provides a high-level overview of class performance without requiring synthesis of each individual submission.

Although the system automatically grouped program-submission pairs by similarity, participants still experienced challenges identifying and utilizing the similarity within each group. In particular, the finer-grained grouping of students based on specific issues introduced significant mental effort. As P15 reflected:
\textit{“At first, I thought, ‘Great! I can answer 30 students’ questions with this one thing.’… But once it broke things down into clusters… each cluster only covered maybe five students’ issues. Then I’d have to switch to another tab just to address two more students’ issues.”}
These challenges were further validated by the NASA-TLX questionnaire results, which revealed consistently high cognitive demands across both systems. Mental and temporal demands were among the highest-rated dimensions (Probe: median = 5.5 and 5.0; Baseline: median = 5.0 and 5.0), with the design probe showing a slightly higher perceived mental load (mean = 4.80 ± 1.93 vs. 4.60 ± 1.26). These findings highlight a need to reduce context switching during feedback review, especially when dealing with large volumes of student submissions.

\begin{figure*}[t]
    \centering
    \includegraphics[width=1\linewidth]{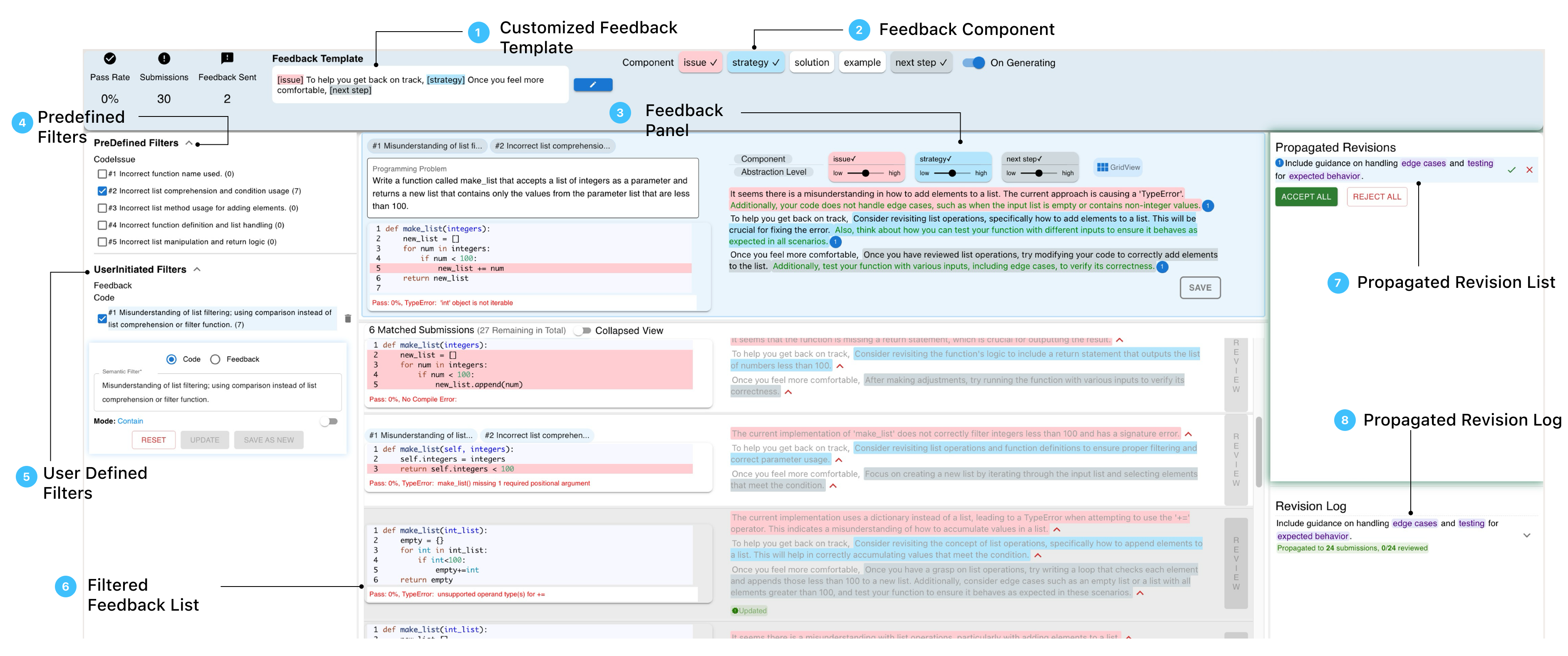}
    \caption{User Interface for \sys{}: \textmd{(1) A customized feedback template is created using (2) modular feedback components like issue and strategy, with (3) feedback panel showing the generated feedback. Users can apply (4) predefined filters or (5) create user-defined filters to target specific errors, which populate the (6) filtered feedback list with matching submissions. Once feedback is applied, (7) a propagated revision list and (8) a revision log track where feedback has been automatically reused and reviewed.}}
    \label{fig:reva-ui}
\end{figure*}
\subsubsection{Instructors employ generalizable revision strategies across similar feedback instances}
\label{sec: pattern3}
Instructors employ a variety of strategies when reviewing similar feedback. Of the 107 valid edits made to the feedback, coded by two researchers, more than half (55) involved changes in abstraction level—that is, modifying the level of detail in the original feedback. The remaining edits were split between content-level edits (41), which introduced new issues not present in the original feedback, and personal-level edits (42), which added encouraging or affirming language for students. Note that one edit can involve multiple strategies. In particular, when feedback is too long, instructors tend to abstract out unnecessary details. As P3 notes, they prefer simple and concise feedback: \textit{“Not only It's very easy for me to understand, and hopefully that's going to help students to quickly understand the feedback.”}. Instructors also employ strategies such as reusing similar feedback for students with similar issues in their code. As P13 suggests \textit{“I’d like to see my own edits somewhere—like a list—so I can reuse specific examples or explanations I’ve given before. It’s not a dictionary, but more like a personal story bank to keep things consistent.”} These activities motivate DG2 in allowing instructors to create reusable components during the feedback authoring process.

\section{Design Goals}
Our formative studies reveal three primary behaviors in reviewing and revising LLM-generated feedback at scale: (1) managing cognitive load during context switching, (2) identifying important content efficiently, and (3) reusing revisions across similar feedback instances. Based on these findings, we came up with three design goals to guide the design of \sys{} for supporting instructors to efficiently validate generated programming feedback at scale.

\begin{itemize}

\item \textbf{DG1. Facilitating a seamless review process based on users' attention.} Instructors preferred reviewing similar issues in batches to reduce context switching, but existing clustering was too coarse-grained (\ref{sec: pattern2}). To address the overload instructors experienced when reviewing the feedback at scale, the navigation process must be efficient and adaptive to users' current focus, allowing seamless context switches among submissions.

\item \textbf{DG2. Users' revision actions should be reusable.} More than half of the edits involved generalizable strategies (abstraction changes, content additions) that instructors wanted to reuse across similar cases (\ref{sec: pattern3}). To avoid compounding instructors' overload with semantically repetitive revisions, there is a need for a system that learns from instructors' revisions and propagates them to applicable submissions, enabling instructors to work on unseen flaws.

\item \textbf{DG3. Enhancing interface support for flexible revision.} Instructors developed implicit mental models for prioritizing content, but current systems didn't align with their intuitive approaches (\ref{sec: pattern1}). To enable users to express their revision intentions in different stages of the review process, the revision action should be able to be initialized efficiently, allowing different levels of feedback modifications.

\end{itemize}
\begin{figure*}[t]
    \centering
    \includegraphics[width=1\linewidth]{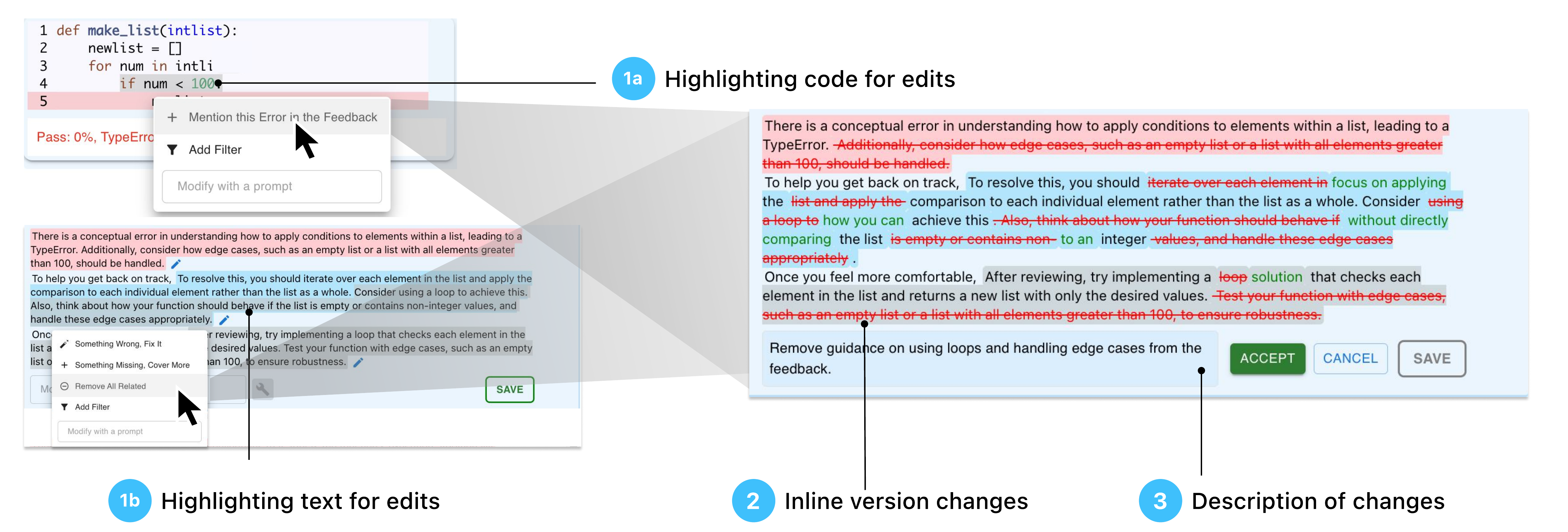}
    \caption{Revisions: \textmd{\sys{} supports different forms of human-AI feedback revision. (1a) Highlighting text and (1b) highlighting code will allow user to edit feedback. (2) Changes are be made inline and (3) a description of changes are displayed}}
    \label{fig:reviseIndividual}
\end{figure*}

\section{\sys{} System}
With the design goals in mind, we designed and developed \sys{}, a human-AI system that streamlines instructors' feedback review and revision experience based on their attention models. \sys{} contains five major panels: (1) a feedback generation panel for component-based feedback generation (Figure ~\ref{fig:reva-ui}.1,~\ref{fig:reva-ui}.2); (2) a feedback review panel that offers flexible and efficient review and revision support (Figure ~\ref{fig:reva-ui}.3); (3) a filter panel that provides pre-defined and user-defined filters for managing the review process (Figure ~\ref{fig:reva-ui}.4, ~\ref{fig:reva-ui}.5); (4) a revision propagation panel that displays the revision actions log and the propagated revisions(Figure ~\ref{fig:reva-ui}.7, Figure ~\ref{fig:reva-ui}.8);  (5) a review list panel that displays the upcoming review in order (Figure ~\ref{fig:reva-ui}.6).

The workflow of \sys{} consists of three key stages. (1) \textbf{Feedback Generation}: User first uploads student code submissions and selects feedback template components (e.g., Issue; Figure~\ref{fig:reva-ui}). To reduce the initial cognitive burden of writing feedback from scratch while maintaining pedagogical structure, REVA generates structured feedback drafts for all submissions using component-based templates examined in the formative study. \sys{} also employs the critical issue detection feature (Figure~\ref{fig:probe}) in the data preprocessing phase to help instructors focus on providing feedback to students with serious issues. (2) \textbf{Attention-Based Review Sequencing}: User highlights problematic code segments/feedback text during review (Figure~\ref{fig:filter}). \sys{} creates semantic filters that automatically reorder the review queue to group similar issues together, which reduces context switching by batching similar problems, allowing instructors to stay focused on one type of misconception at a time. (3) \textbf{Revision Propagation and Validation}: User makes targeted revisions to feedback (Figure~\ref{fig:reviseIndividual}) and \sys{} automatically applies similar revisions to matching submissions, waiting for individual decision (Figure~\ref{fig:propagation}). This design aims to reduce repetitive editing while preserving instructor control over every final revision.

The section begins by demonstrating an example user scenario of \sys{}, detailing its key interfaces and features, their rationale and how they connect to the design goals, and concludes with a concise overview of its implementation.

\subsection{Example User Scenario}
Bob is working on providing students with programming feedback on a coding assignment in his introductory-level programming class, which has hundreds of students. He aims to give comprehensive feedback to students with critical issues. However, after a quick screening, Bob still finds the workload of manually writing it all unacceptable. He decides to use \sys{}.

\begin{figure*}[t]
    \centering
    \includegraphics[width=1\linewidth]{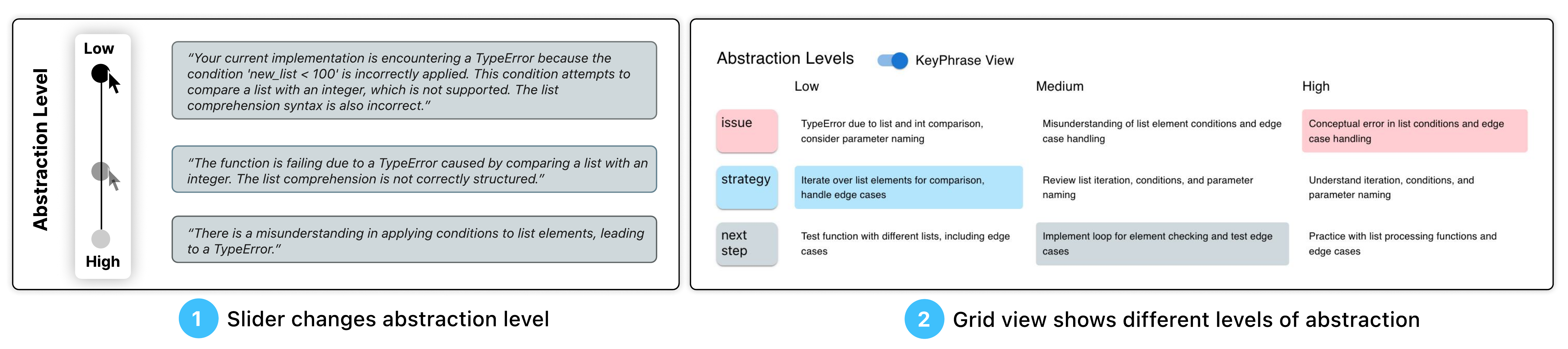}
    \caption{Abstraction Levels: \textmd{Feedback content can be controlled by the level of abstraction of the content through a (1) Slider and (2) Grid that shows all feedback combinations.}}
    \label{fig:abstractionLevel}
\end{figure*}

Bob uploads students' code submissions into \sys{}, where he begins by setting up the \texttt{feedback template} with desired \texttt{components} (Figure~\ref{fig:reva-ui}.1; \ref{fig:reva-ui}.2). After \sys{} finishes the feedback generation, Bob starts the feedback validation process. For each code-feedback pair, he needs to make sense of both the code and feedback and cross-check them to ensure consistency and effectiveness. During the review process, he recognizes a misconception in a student's code that can be very common. To see more and keep an eye on this type of issue, he initializes a \texttt{semantic filter} by selecting the lines of code he finds the misconception (Figure~\ref{fig:filter}.1). 
Then he starts to read the generated feedback and realizes that this issue is not mentioned. He initiates the revision in situ by directly selecting the code to include the issue in the feedback (Figure~\ref{fig:reviseIndividual}.1a). After doing so, he also adjusts the \texttt{abstraction level} of components to encourage more self-reflection on the student's end (Figure~\ref{fig:abstractionLevel}).

After his revision, Bob notices that \sys{} has extracted the meaning of the modifications and propagated it to many other code-feedback pairs (Figure~\ref{fig:propagation}.2). When he proceeds to the next submission, he finds that \sys{} pulls up a similar submission based on the active filters and his revision, and adapts the revision to the current feedback for him, waiting for verification (Figure~\ref{fig:propagation}.4). After a quick check regarding the code and revision, Bob accepts the propagation and works on verifying the other parts of this feedback and looking for new misconceptions that are not recognized by \sys{}. In rounds of review and revision process, \sys{} learns contents caught Bob's attention, diverse types of misconceptions, and Bob's feedback style, adapting itself based on Bob's validation process. Bob is now able to review the feedback with lower context-switch and navigation effort, and instead of repetitively initiating similar revisions, he can spend more time working on unseen flaws.

\subsection{Interface \& Features}
This section describes the following key design features of \sys{} to support users' feedback review and revision process with user attention-based adaptation.

\subsubsection{Component-based feedback generation}
To enable instructors to personalize the feedback generation and efficiently locate different high-level segments in the feedback in the review process (DG1), we incorporate the component-based feedback generation process and the corresponding highlightings in the probe system.

Instead of generating the template each time, users use a single template to steer the feedback generation in \sys{}. To generate pedagogically comprehensive feedback that maximum feedback's impact, five feedback components (\colorbox[HTML]{fec6cc}{Issue}, \colorbox[HTML]{abe2fa}{Strategy}, \colorbox[HTML]{d8eac3}{Solution}, \colorbox[HTML]{ffe8ae}{Example}, \colorbox[HTML]{c9d3d7}{Next step}) are contained in \sys{}. These five components are derived from Hattie et al.'s feedback model ~\cite{hattie2007power}, which states that effective feedback answers three questions: \textit{(1) What are the goals? (2) What has been done to progress towards the goals? and (3) What activities need to be undertaken to make better progress?} To facilitate easy verification of feedback, we incorporated highlighting as a way for instructors to review how each component is incorporated into the final feedback, as well as the related code or conversation that reflects the feedback.

\subsubsection{Individual Feedback Revision}
To facilitate efficient revision on misaligned feedback, \sys{} offers different forms of revision interactions in the review panel. Besides manually typing, to allow users to initiate the revision most directly and flexibly (DG3), drawing on previous work on human-AI text editing~\cite{ExecutableandVerifiableText-Editing}, \sys{} provides users with human-AI revision support from different levels, i.e., content, abstraction, and personal levels. Content-level revisions address the issue of low recall and low precision regarding the coverage of misconceptions of the code in the feedback, while abstraction level revisions align the feedback with the severity of the student's issue based on the instructor's domain knowledge, aiming for maximizing the learning outcome. The personal-level revisions focus on the instructor's personal language (e.g., tune and grammar) and mentor style, providing students with personal level greetings and encouragement.

\sys{} also enables users to suggest feedback revision with different levels of indication, i.e., explicit and implicit, located in diverse types of content, i.e., programming problem, code, and feedback. Users can ask for feedback revision using the general revision query component at the bottom of the review panel. 

As users' revision intention may be grown from different content (e.g., programming problem, code, feedback) during review, \sys{} also allows them to suggest a revision in situ (Figure~\ref{fig:reviseIndividual}.1a, 1b). For example, after reading a code without a return statement, the user checks the corresponding requirements in the programming problem, and they can directly select the text "return a new list" to suggest a revision. Based on our formative study's observation of instructors' revision strategies, we designed multiple shortcuts for in-situ revision for users to express their intention implicitly to lower the typing effort (Figure~\ref{fig:reviseIndividual}.1). A shortcut for mentioning issues applies to in-situ revision from the programming problem and code. Three shortcuts about fixing errors in feedback, expanding feedback, and removing feedback were applied to the in-situ revision from the feedback. The in-situ revision query takes the whole code-feedback pair as the context and emphasizes the users' selection to establish the revision.

After the user submits a revision query and receives the response from the system, the revision will be presented with inline version changes (Figure~\ref{fig:reviseIndividual}.2) and a description of changes (Figure~\ref{fig:reviseIndividual}.3), and the user can choose to accept or dismiss the revision suggestion after the evaluation.
\begin{figure*}[t]
    \centering
    \includegraphics[width=1\linewidth]{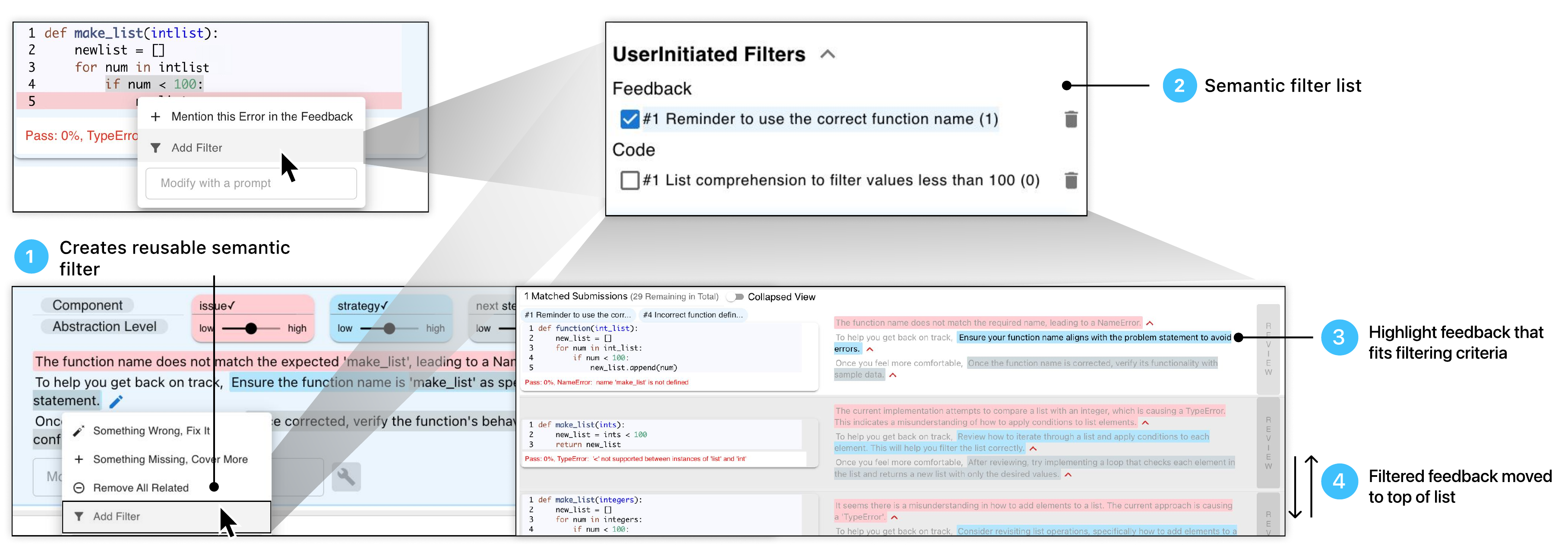}
    \caption{Filtering: \textmd{Filtering allows instructor to focus on specific issues in student submissions. (1) Filters are initiated by directly highlighting related comment or code. (2) After clicking "Add Filter", a one line summary of the filter is created by LLM. Upcoming submissions are (3) highlighted and (4) reordered such that submissions that fits filter requirement will be reviewed first.}}
    \label{fig:filter}
\end{figure*}

\subsubsection{Abstraction-level control}
To facilitate flexible and efficient revision of feedback (DG3), drawing on Soylent’s Shortn component~\cite{SoylentBernstein} that allows slider-based content adjustments, \sys{} offers direct control of the abstraction levels of each feedback component (DG3, Figure~\ref{fig:abstractionLevel}).
While the in-situ and general revision queries support all three levels of edits we identified in the formative study, the abstraction level of feedback is the only one-dimension level as there are only \textit{Abstract} v.s. \textit{Specific}. Based on the our observed feedback variation on the abstraction level in the formative study and the prior work on text semantic levels~\cite{ExtractiveAbstractiveWorledge, FourLevelModelLundgard}, we designed a three-level abstraction control for feedback alignment, where the low abstraction level tends to be extractive and the high level tends to be conceptual.

The user can use the slider at the top of the review panel to adjust the abstraction level for a specific component based on their understanding of the student's misconceptions and their domain knowledge, such as overall class progress. The user can open the grid view (Figure~\ref{fig:abstractionLevel}.2) to learn the differences by viewing all abstraction levels simultaneously. We also maintain the consistency of content-level information (i.e., issue coverage and correctness) across different abstraction levels in the generation process to avoid any potential conflicts that influence the review process. 

\subsubsection{Filters}
To reduce users' cognitive switching costs by sequencing and rendering reviews in the way aligned to users' attention (DG1), \sys{} employs both pre-defined and user-initiated semantic filters.
Unlike public grading tools like Gradescope which uses similarity-based clustering, \sys{} employs both a pre-defined filter based on code embedding similarity with general issue summarization, and a semantic filter that adapts to the user's nuanced implicit attention models in review.

There are two types of semantic filters (i.e., code and feedback, Figure~\ref{fig:filter}.2) that capture users' attention model at the semantic level, and emphasizes their focus in individual reviews and reduce cognitive switching costs in the overall review process. The semantic meaning of the filter also serves as a feedforward~\cite{min2025feedforward}, allowing users to understand and align the filter to their mental model before evaluating the outcome.

The user can create a semantic filter by defining it from scratch in the filter panel. \sys{} also allows users to initiate a semantic filter based on an instance during the review process. For example, if the instructor finds a code segment "\textit{new\_list+=num}" problematic and wants to focus more on this in the review process, they can select the code segment and click "Add Filter." \sys{} will interpret the instructor's intention based on the entire context of the code-feedback pair and add a code semantic filter of "Misconception about list concatenation." After that, the review queue will be reordered based on the matching conditions of active filters and code similarity. In each pair, the matched code's corresponding lines will be highlighted for code filters to guide the user's attention. For the feedback filter, matched keywords and key phrases will be rendered in a darker color for better skim performance based on previous work~\cite{AnAI-ResilientTextZiweiGu}. After creating a semantic filter, the user can make further alignments on it in the filter panel.

For semantic filter creation, we use an off-the-shelf LLM to interpret the particular semantic meaning of the user-selected content when initiating it through instance. For the semantic match, we directly provide the corresponding content with the context of code-feedback pairs to have the matching results.

\subsubsection{Revision Propagation}

To facilitate attention-based review (DG1) and reusable revision actions (DG2), \sys{} offers revision propagation and corresponding review order adaptation.
Taking inspiration from previous research about reusing and propagating users' actions~\cite{head2017writing, WranglerKandel2011, PropagatingKhan2022}, \sys{} supports instructors by learning their revision actions and propagating to upcoming applicable submissions while adapting their order of review. In the formative study, instructors spent their major efforts in identifying patterns and initiating revision manually, if necessary. In this process, the semantic repetition of both the information they reviewed and the content they produced in the revision was underutilized due to the misalignment between their practice of review order and the order that adapted to their attention model.

Thus, through leveraging the explicit attention claimed by revision actions, we empower instructors by automatically prioritizing the corresponding code-feedback pairs in the upcoming revision, as well as suggesting similar revisions based on the original revision and the current context, allowing them to shift their efforts to discovering unseen flaws and addressing them accordingly.

\begin{figure*}[t]
    \centering
    \includegraphics[width=1\linewidth]{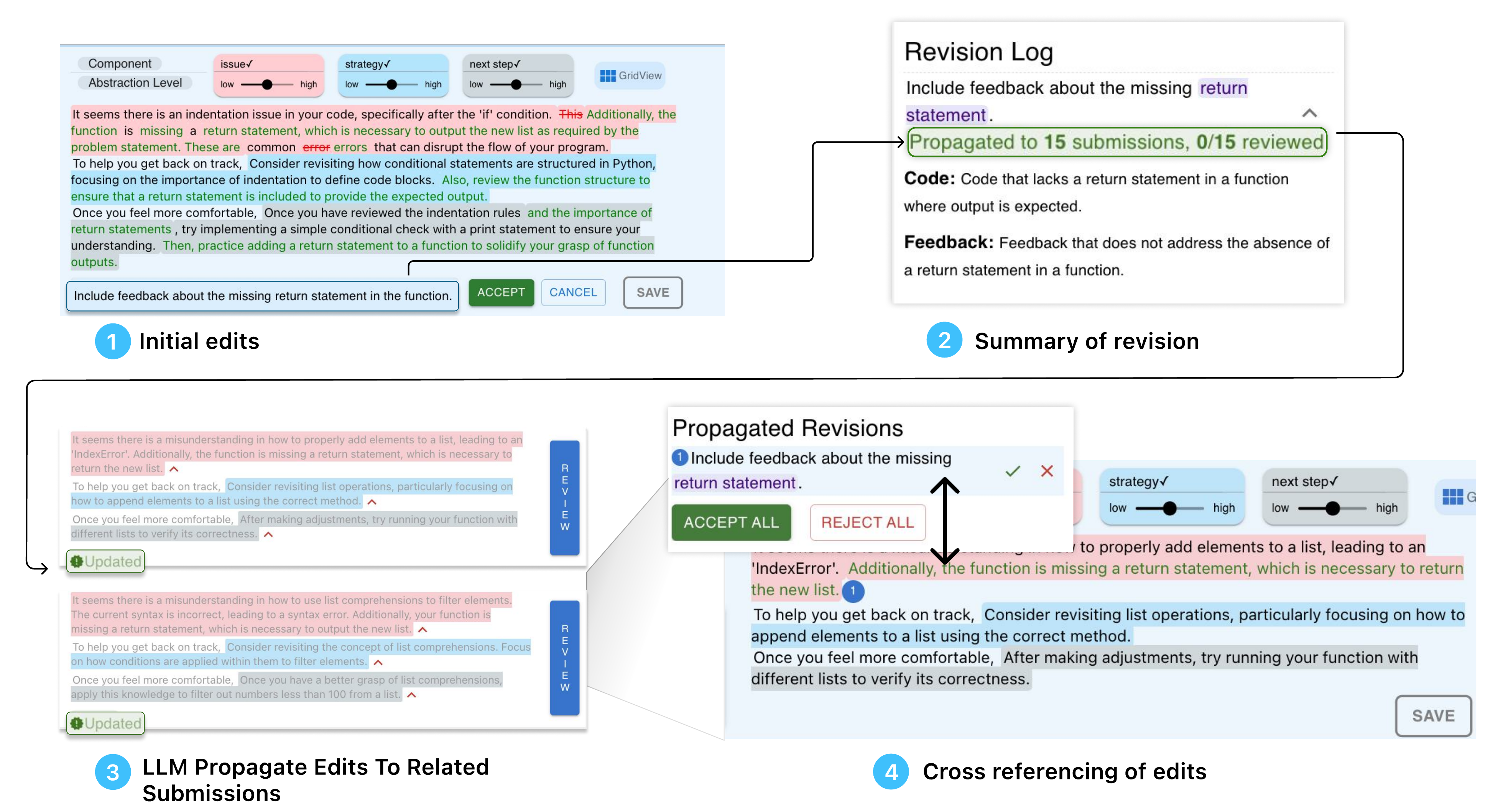}
    \caption{Feedback propagation: \textmd{Feedback propagation automatically propagate revision to similar feedback. After the user (1) accepts a revision suggestion, (2) \sys{} extracts the revision with code and feedback patterns, and (3) propagates the revision to applicable feedback. When the user reviews a propagated code-feedback pair, (4) the propagated revision and its' corresponding edits would be visually connected.}}
    \label{fig:propagation}
\end{figure*}

After the user initiates a revision and accepts the suggestion, \sys{} extracts the action of the feedback revision with the corresponding code and feedback patterns. Next, it will match this action with appropriate submissions and suggest their revision based on the revision action, code pattern, and feedback pattern (Figure~\ref{fig:propagation}.2). \sys{} also reorders the review queue based on the propagation result (Figure~\ref{fig:propagation}.3) to enable the instructor to pay attention to their focus of revisions easily in the review process.

When the instructor starts to review a code-feedback pair that has been propagated with previous revisions (Figure~\ref{fig:propagation}.4), the propagated revisions will be listed on the propagation panel. The corresponding revision will be presented with inline version changes. For navigation, each propagation has an ordinal number in both the propagation panel and the feedback panel, and the instructor can click to focus on one propagation. After verification, the instructor can choose to accept or reject the propagation revisions.

To achieve revision propagation. We use an off-the-shelf LLM to first understand the high-level revision goal, and extract the patterns (e.g., code pattern and feedback pattern) that specifically correspond to the revision. After that, we collect all upcoming code-feedback pairs that match both patterns and then apply the high-level revision goal to the pairs. We also maintain the coherency of all propagated revisions to avoid conflicts and repetition on a single code-feedback pair.

\subsection{Implementation}
\sys{} was implemented as a web-based programming feedback creation and validation tool using React, FastAPI for its core functionality, and OpenAI's GPT-4o\footnote{\url{https://platform.openai.com/docs/models/gpt-4o}} and text-embedding-3-large\footnote{\url{https://platform.openai.com/docs/models/text-embedding-3-large}} for specific features. We used Firebase\footnote{\url{https://firebase.google.com/}} to track and record participant’s interaction log data.

The prompts used in the system were iteratively engineered to map directly onto the cognitive load and reuse challenges \sys{} targets, then empirically stress-tested (response time $\approx3$ seconds, error rate < 5\% on trial). Hallucination risk is further bounded by template constraints and mandatory human verification.
We also adopted well-known prompting engineering techniques. To gain responses from the perspective of an instructor, we used persona settings~\cite{10.1145/3586183.3606763}. To improve the performance on complex and long-context tasks, we adopted AI chains~\cite{10.1145/3491102.3517582} and few-shot prompts~\cite{NEURIPS2020_1457c0d6}. Meanwhile, to achieve a shorter response time, our prompting pipeline was designed to execute tasks in parallel.

\section{Lab study with instructors}
We conducted an in-person within-subject user study to examine \sys{}'s usability and effectiveness for supporting instructors' review and revision process of LLM-generated feedback.
Our study is approved by the IRB at our institute.

\subsection{Participants}

We recruited 12 participants (4 female, 8 male) who have both teaching and programming experiences at a four-year university via personal networks, local mailing lists, and snowball sampling. During the study, participants were asked to validate and send feedback to students based on their codes that were collected in a large programming course at our institution. Each participant was compensated with \$30 USD for their time and effort.
\subsection{Protocol}
\subsubsection{Datasets}
To ensure the authenticity of the data participants interacted with, we collected real data from three large introductory-level university programming course coding exercise sessions. The Python problems were about writing a function for \textit{filtering}, \textit{counting}, and \textit{summation} regarding a list, accordingly. We used the \textit{filtering} and \textit{counting} sessions for formal study, and the \textit{summation} for the tutorial and warm-up session. 
\begin{table*}[t]
\centering
\small
\begin{tabular}{l c c c c c c c c}
\toprule
\textbf{Condition} & \textbf{Mental demand} & \textbf{Physical demand} & \textbf{Temporal demand} & \textbf{Performance} & \textbf{Effort} & \textbf{Frustration} &\textbf{Aggregated} \\ 
\midrule
\textbf{Baseline} 
& 5.0 (4.50 ± 1.31) 
& 2.0 (2.58 ± 1.56) 
& 3.5 (3.83 ± 1.70) 
& 2.0 (2.75 ± 1.54) 
& 4.0 (3.83 ± 1.64) 
& 3.0 (3.17 ± 1.80) 
& 3.3 (3.44 ± 1.15)
\\ 
\textbf{\sys{}} 
& 3.0 (3.58 ± 1.62) 
& 1.5 (1.92 ± 1.44) 
& 3.0 (3.17 ± 1.64) 
& 2.0 (2.00 ± 0.95) 
& 4.0 (3.50 ± 1.57) 
& 2.0 (2.67 ± 1.37)
& 2.7 (2.81 ± 1.07)
\\
\bottomrule
\end{tabular}
\caption{Response to NASA TLX items. Format: median (mean ± standard deviation). We also report the aggregated score as all items together culminate to the latent variable task load ~\cite{10.1145/3582272, babaei2025should}. }
\label{tbl:nasa-tlx}
\end{table*}
\subsubsection{Conditions}
We used a within-subject study design where each participant used the following two systems:

\begin{itemize}
    \item \textbf{Baseline:} a baseline version of \sys{} without the \texttt{semantic filter} and the \texttt{revision propagation}. The system still contained the pre-defined filters and all individual revision features, and shared the same feedback generation process with full \sys{}.
    \item \textbf{\sys{}:} a full version of \sys{} with all its features.
\end{itemize}

\subsubsection{Task}
Participants shared the same form of task in each condition. They were asked to provide 30 validated programming feedback on incorrect students' code submissions from one coding exercise session as an introductory-level programming instructor using the system for 20 minutes.

\subsubsection{Study Procedure}
Each study was conducted in person in a lab setting and lasted around 75 minutes. At the beginning of each study session, we collected informed consent from the participants after introducing the goal and the process of the study. After that, we gave participants an explanation of the context of the data and the task used in the study. For each condition, following a general introduction and explanation, we offered a detailed tutorial of the system and then asked the participant to explore the system on the trial dataset to warm them up and answer their questions. Once they were familiar with the system, participants were asked to complete the feedback validation task. At the end of each condition, participants completed a survey with Likert scale questions (including a NASA-TLX questionnaire) and participated in a semi-structured interview. All studies were screen- and audio-recorded and participants were asked to think aloud while completing the survey questions. The orders of conditions and datasets were counterbalanced and randomized.

\section{Results}
By recording the feedback participants generated, revised, and sent in the user study, we collected 475 validated feedback messages in total, along with 840 revisions made by participants. To evaluate the quality of the feedback, two researchers coded the feedback quality regarding the issue coverage and overall comprehensiveness~\cite{10.1145/3359174}. The annotation process was as follows: First, we sampled 89 out of 475 (Around 18\% coverage) code-feedback pairs from each participant and each condition proportionally and randomly shuffled the sampled data. After collecting all the unique codes, we annotated misconception types for each code, and used it as the ground truth for issue coverage. For each validated code-feedback pair, we annotated the covered misconceptions in the programming feedback, and coded the overall feedback quality to three levels of score, i.e., $-1$ for \textit{incorrect feedback}, $0$ for \textit{shallow feedback}, 1 for \textit{comprehensive feedback}. We used T-test for temporal, count, and proportion data.

On average, participants validated 20.67 ($\sigma=9.63$) feedback using \sys{}, and 18.92 ($\sigma=10$) feedback using Baseline. For each code-feedback pair, it took them $11.14\%$ less time to review using \sys{} ($\mu=79.73~seconds, \sigma=59.47$) than Baseline ($\mu=88.85~seconds,  \sigma=64.55$). Instructors also completed significantly more revisions using \sys{} ($\mu=44.00, \sigma=16.36$) than Baseline ($\mu=26.00, \sigma=6.70; p<0.01$).

Table~\ref{tbl:nasa-tlx} shows the NASA-TLX survey result. While participants using \sys{} perceived low levels of both mental ($Median=3.0, \mu=3.58, \sigma=1.62$) and physical demands ($Median=1.5, \mu=1.92, \sigma=1.44$), quality-wise, as shown in Table~\ref{tbl:feedback quality}, they also produced feedback with significantly higher recall (\sys{}: $\mu=0.86, \sigma=0.20$, Baseline: $\mu=0.55, \sigma=0.25$, $p<0.0001$) and precision (\sys{}: $\mu=0.90, \sigma=0.20$, Baseline: $\mu=0.71, \sigma=0.28$, $p<0.001$) regarding misconception coverage, as well as overall feedback quality (\sys{}: $\mu=0.34, \sigma=0.48$, Baseline: $\mu=0.04, \sigma=0.42$, $p=0.003$) than the Baseline.

To reflect the cost of context-switching~\cite{wylie2000task}, we define the reaction time in the study as the amount of time participants spend before performing the first concrete action using the system (e.g., initiating revision) when validating a new feedback pair. We found that participants using REVA had significantly lower reaction time ($\mu=18.69~seconds, \sigma=8.10$) than the Baseline ($\mu=29.77~seconds, \sigma=16.32$, $p<0.05$).

\subsection{\sys{} creates a flexible feedback authoring workflow for Instructors }
\subsubsection{Filters help instructors make sense of different issues faced by students }
9 of 12 participants incorporated semantic filters in their validation workflow and each of them created 3.67 ($\sigma=1.00$) filters on average.
Participants noted that semantic filters help them make sense of and prioritize student code and the issues present in submissions. P2 appreciated that filters shorten the process of determining which student submission to review first. They explained: \textit{“There were some misconceptions about using built-in types in Python. If you identify this kind of issue, you can edit it into a filter. The system will then create a filter with a corresponding semantic meaning. Your review order can be based on this filter, and the system will highlight the parts of the code that match it. This can help accelerate your understanding of students’ code. ”} Similarly, P1 emphasized how focusing on one type of issue helps reduce cognitive load and improve grading efficiency: \textit{“I really like being able to highlight the system’s suggestions, add to the identified errors, and create filters. Especially in the beginning, it really helps if I just want to focus on one issue. That way, I don’t have to switch my attention to different types of errors — I can stay focused on one and grade more efficiently.”}

Interestingly, several participants (P1, P2, P3, P5, P6, P10) felt that predefined filters better supported them in reviewing feedback—especially when they were new to the system. P1 and P6 noted that predefined filters provide guidance on what to focus on when starting out. As P1 said, \textit{“Sometimes [I’m] not sure like what I should [filter] in this case, but in here, okay, I can see clearly like what are some predefined filters and I can use that.”} They suggested that predefined filters help avoid decision fatigue and can scaffold the process of identifying key issues. P2 added that predefined filters can nicely complement propagated revisions, describing them as a “cherry on [the] cake” in supporting effective feedback review.
\subsubsection{Feedback Generation helps reduce cognitive workload during feedback authoring (DG3)}
Participants felt that the system's support for feedback generation significantly reduces the cognitive load of synthesizing multiple sources when authoring feedback. P5 appreciated the large language model’s ability to describe student issues concisely, providing a useful starting point: \textit{“Compared to writing the feedback from scratch, which is very demanding, I think the AI helps because the basic issue is like the logic error in the for loop, which it very nicely describes—like, that's the logic and they need to revisit it.”}

Participants also valued the ability to edit feedback by directly referencing code or other relevant sources. Each participant initiated 19.08 ($\sigma=13.34$) revision suggestions in situ on average in the whole study. P6 noted how intuitive the interaction felt, describing it as \textit{"straightforward"} and liked the ability to select area of interests when prompting the AI. P7 contrasted this streamlined process with their usual grading workflow, which often involved juggling multiple tools and windows: \textit{“I think that was really nice. Because that’s the thing—like sometimes when I’m grading, I have to copy [the code] and put that in a Google Doc or something, and then tell them ‘this was the line that wasn’t working.’ So here, I can just highlight it—it goes right to the feedback.”} 
\begin{table}[b]
    \centering
    \small
    \begin{tabular}{l c c c c c}
    \toprule
    \multirow{2}{*}{Condition} & &\multicolumn{2}{c}{Misconception Coverage} & &\multirow{2}{*}{Quality} \\
    \cline{3-4}
    & & Precision & Recall &  \\
    \midrule
    Baseline & & 0.71 & 0.55 & & 0.04 \\
    \sys{} & & \textbf{0.90}$^{**}$ & \textbf{0.86}$^{***}$ & & \textbf{0.34}$^{*}$ \\
    \bottomrule
    \end{tabular}
    \caption{Precision and recall of code misconception coverage in the feedback, and the overall feedback quality score (range from -1 to 1) in the sampled dataset. $*$ indicates $p < 0.01$, while $**$ indicates $p < 0.001$, and $***$ indicates $p < 0.0001$}
    \label{tbl:feedback quality}
\end{table}

\subsubsection{Instructor changes abstraction level based on student's need}
When authoring feedback, participants adjusted the level of abstraction to match the needs of the student and the nature of the issue being addressed. This allowed them to control the level of detail, tone, and clarity depending on what they thought would be most helpful. P6 explained: \textit{“Having different abstraction levels are helpful—like how detailed you want the feedback. Sometimes students don’t really like to read too much… so it’s nice to have a straightforward quick answer, and if you want to, you can make it more detailed.”} Similarly, P10 described shifting abstraction levels based on the type of issue they were addressing.

\subsection{\sys{} reduces mental demand for instructor while maintaining feedback quality. }
\subsubsection{Feedback propagation reduces repetitive work from instructors (DG2)}

Participants discussed feedback propagation helps them offload repetitive work of writing feedback to students with similar issues. This allows them to focus on other efforts.  Specifically, instructors noted that using feedback propagation allowed them address recurring errors found in multiple submissions. P3 suggests that \textit{"It gets annoying when you have to repeat "Missing return statement" on 10 people back to back,...,I know from  TAing and teaching from Python that like a lot of people might have similar issues, and I feel like I'm a broken record like missing return statement, missing return statement. So having that be something that was  automatically propagated was really handy"}. Other participants such as P4 like the benefit of not needing to repeat work of looking into all similar submission before sending out feedback. P4 states that propagation \textit{helps me address overlapping errors (from students) easily because I don't have to go through each code entirely to see those errors. ... I don't have to select the error go through the process again, and I can just accept the propagation 
revision}. 
Because of the reduce in redundancy in work, instructors explain that it helped them focus on other important tasks. For instance, P4 explained that reducing redundant work of addressing repeated code error allowed them to redirect their effort to find new errors in students. P11 also states the benefit of propagation as a way to remind them of previously identified issue. \textit{"If I identify the issue by myself—maybe I noticed the same problem in a previous submission, like with a built-in type such as list or int—I might forget it in the next one. Propagation is a good way to remind me of issues I previously identified"} (P11).
\begin{figure*}[t]
    \centering
    \includegraphics[width=1\linewidth]{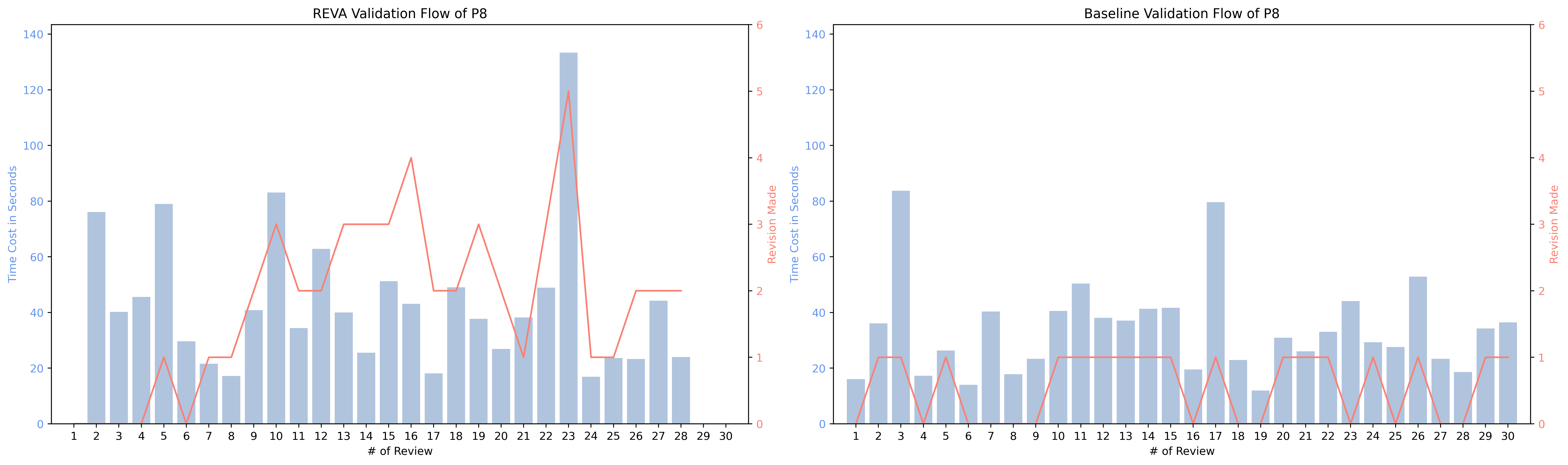}
    \caption{Number of revisions (orange) completed by P8 through revision suggestion with the cost of time (blue) for each code-feedback pair's validation. Y-axis stands for time/count, and X-axis represents the number of reviewed code-feedback pairs.}
    \label{fig:P8Flow}
\end{figure*}
\subsubsection{Feedback propagation helped instructor to maintain higher feedback quality (DG2)}
As shown in Table~\ref{tbl:feedback quality}, feedback validated using \sys{} established significantly better misconception coverage and higher quality.
Feedback Propagation not only offloads repeated work from instructors, but also allows authoring of consistent and personalized feedback that can be sent to different students. Specifically, P1 suggests that feedback propagation \textit{"adapts to my style of feedback so that I don't have to keep adjusting every time (I send out feedback)"}. P12 thought feedback propagation can maintain the quality of feedback because they can reuse previous editing prompts. \textit{" I don't need to manually add prompts—I can just reuse the ones from previous experience" } (P12).

\subsubsection{Feedback propagation gives participants 
sense of control}
Despite the automated and non-deterministic nature of feedback generation, participants perceived high level of control when using \sys{} ($\mu = 5.83$, $\sigma = 1.337$). During the post study interview, 10 out of 12 participants reported high levels of control. Participants shared myriad of reasons that contributed to their sense of control. For instance, P11 explains that despite having slightly less control on how the feedback is phrased, they feel like they can have "control over the content of the feedback". Others such as P3 perceives high control as having final decision on what to include and not include in the prompt:  \textit{"the ability to type in the prompt and reject what ChatGPT said if you didn't like it"}. However, some participants also found the features offered by \sys{} limiting, lowering their perception of control on the system. For instance, P7 perceives the reliance of AI in generating their feedback content and edits as a perception of low control. \textit{"I wasn’t controlling it a lot, the AI was. I mostly just prompted the AI ... So I wasn’t really the one driving the interaction, at least not the writing part"} (P7).

\section{Discussion}
In this section, we attempt to explain the reasons that led to our findings and discuss the limitations of our work.
\subsection{\sys{}'s participants are more engaged in feedback validation across time}
While participants' perceived effort (Table~\ref{tbl:nasa-tlx}) for \sys{} and Baseline is on par, with semantic filter and revision propagation, they made significantly more revisions and produced better feedback during the study session. Though participants in both conditions shared the common validation behavior at the beginning of the review, from our observation, participants (P1, P2, P5, P7, P8, P9), using \sys{} tended to be more consistent in their revision engagement than the Baseline. For instance, in Figure 8, when using \sys{}, P8 consistently made revisions through evaluating and accepting the suggestion, but when using Baseline, the participant dropped out of the efficient revision workflow early. As participants also claimed a high level of control while using \sys{}'s revision propagation feature, we anticipate that \sys{}'s user attention-based alignment also enables users to engage in the feedback validation process with lower effort and, thus, longer time.

\subsection{Human-AI Co-adaptation for Feedback Revision}
As \sys{} continued to learn from users' attention, we also observed a reciprocal adaptation in participatns' validation behavior when interacting with propagated revisions. Without the revision propagation feature, participants followed a predictable pattern of validating code-feedback pairs through direct sensemaking of both elements. However, when presented with propagated revisions, their cognitive approach shifted. Rather than treating each code-feedback pair as an isolated unit requiring full assessment, participants like P11 began using propagated revisions as a cognitive anchors to guide their attention to potential misconceptions in code and misalignment in feedback. 

\subsection{Human-AI Collaboration in Large-scale Validation Tasks}
\sys{} and our study contributes several insights about human-AI collaboration in large-scale validation tasks. Firstly, using highlighting/selection as implicit signals of user priorities can automatically adapt AI workflows without explicit configuration. This is applicable wherever humans review AI-generated content (e.g., students' essays grading, legal documents) at scale. Secondly, the pattern of extracting user intent from accepted edits and applying similar changes to matching instances generalizes to any domain with semantic repetition (e.g., patent review, medical chart analysis). Lastly, reordering review queues based on user focus patterns reduces context switching costs. Any task involving sequential validation of similar items could benefit from this.

\subsection{Limitations}
Our focus was on how \sys{} affects instructors' ability to provide feedback that they consider comprehensive and effective. However, we did not investigate the impact on students, and our sample size is limited. Future research will examine how LLM-generated feedback, validated by instructors, influences both the perceived and actual outcomes for students in the medium and long term. Additionally, we aim to explore how this feedback can be relayed back to instructors to offer them deeper insights and facilitate more personalized teaching approaches.

Although we used authentic student data, our evaluation was conducted in a lab setting rather than in a genuine grading environment. A reduced set of pressures (i.e., the cognitive demands of misleading real students) may impact the available attention that the instructor has to review and revise the feedback. Additionally, there is a discrepancy in the length of the review process; our feedback review lasted only 20 minutes for each condition to prevent potential bias due to fatigue, whereas a typical real-world grading session can span several hours. Nevertheless, given that \sys{} demonstrates its potential to keep instructors engaged during the feedback validation process, we contend that the overall net effect of \sys{} in real grading scenarios could be even more beneficial.

\sys{} uses a user attention-based adaptation approach to align the review sequence with users' mental models. However, it does not incorporate any state-of-the-art task sequencing algorithms to optimize the global review sequence. In reflection on the study, some participants experienced high mental demand at the beginning of the task due to the initial code-feedback pairs containing more extensive misconceptions than average, making them particularly challenging to review.

Though \sys{} augments users' ability to validate AI outputs, helping them lower the risk of hallucination (Table~\ref{tbl:feedback quality}; REVA helped participants gain both higher recall and precision), LLMs are still prone to hallucination, lack awareness of learning context, and cannot replicate the nuanced pedagogical judgment and adaptive delivery that effective educators provide. Future work could focus on combining techniques that produce deterministic outputs (e.g., programming language techniques) with LLMs to tackle the issue from the system aspect.

\section{Conclusion}
In this paper, we introduced \sys{}, a human-AI system designed to enhance instructors' capacity to efficiently review and refine large volumes of AI-generated programming feedback through user attention-based adaptation. 
By integrating adaptive content sequencing and revision propagation, \sys{} capitalizes on the semantic patterns inherent in programming feedback, thereby minimizing cognitive load from frequent context shifts and reducing redundant revision efforts during feedback review.
Our within-subject study with programming instructors validated that \sys{} significantly improves both the efficiency and quality of feedback review and revision compared to the baseline system. Specifically, instructors generated higher-quality, more precise, and more comprehensive feedback without incurring significant increases in validation time.
These results underscore the potential of adaptive human-AI collaboration to address the persistent challenge of scaling personalized feedback in programming education. 
By effectively leveraging patterns of instructor attention and intention, \sys{} offers a promising blueprint for designing future educational tools capable of delivering high-quality, personalized feedback at scale, thereby transforming instructor workloads into cognitively sustainable and pedagogically rewarding experiences.


\bibliographystyle{ACM-Reference-Format}
\bibliography{ref}


\end{document}